\documentclass[10pt]{iopart}
\usepackage{CJK}
\usepackage{graphicx}

\begin{document}
\begin{CJK*}{GBK}{song}

\title[DBT based on CRDS]{Doppler broadening thermometry based on cavity ring-down spectroscopy
       } %


\author{C.-F. Cheng, J. Wang, Y. R. Sun, Y. Tan, P. Kang, S.-M. Hu}

\address{
Hefei National Laboratory for Physical Sciences at Microscale, $i$Chem center,
    University of Science and Technology of China, Hefei, 230026 China
}%
\ead{smhu@ustc.edu.cn}


\begin{abstract}
A Doppler broadening thermometry (DBT) instrument is built based on cavity ring-down spectroscopy (CRDS)
 for precise determination of the Boltzmann constant.
Compared with conventional direct absorption methods,
 the high-sensitivity of CRDS allows to reach a satisfied precision at lower sample pressures,
 which reduces the influence due to collisions.
By recording the spectrum of C$_2$H$_2$ at 787~nm,
 we demonstrate a statistical uncertainty of 6~ppm (part per million) in the determined linewidth values
 by several hours' measurement at a sample pressure of 1.5~Pa.
As for the spectroscopy-determined temperatures,
 although with a reproducibility better than 10~ppm,
 we found a systematic deviation of about 800~ppm, which is
 attributed to ``hidden'' weak lines overlapped with the selected transition at 787~nm.
Our analysis indicates that it is feasible to pursue a DBT measurement toward the
 1~ppm precision using cavity ring-down spectroscopy of a CO line at 1.57~$mu$m.

\end{abstract}

\pacs{  33.20.Ea, 31.30.J-,
    }
%
%
%
\ioptwocol

\section{Introduction}

The kelvin unit will be redefined on an exact value of the
 Boltzmann constant $k_B$~\cite{Mills2006Metrologia_Kelvin},
 which directly relates the thermodynamic temperature to thermal energy.
The present CODATA~\cite{CODATA2010RMP} recommended value of $k_B$ is
 $1.3806488(13)\times 10^{-23}$~J/K,
 inferred from a group of results obtained from the acoustic gas thermometry
 (AGT)~\cite{Colclough1979_AGT_NPL, Moldover1988PRL_AGT_NIST, Pitre2009CRP_AGT_LNE, Sutton2010IJT_AGT_NPL, Gavioso2010Metrologia_AGT_INRIM, Pitre2011IJT_AGT_LNE},
 the refractive index gas thermometry (RIGT)~\cite{Schmidt2007PRL_RIGT_NIST},
 and the Johnson noise thermometry (JNT)~\cite{Benz2011Metrologia_JNT_NIST}.
The combined relative uncertainty of $k_B$ is 0.91~ppm.
The contributions from AGT results in present $k_B$ value are dominant
 since their uncertainties are much smaller than those from other methods.
A value with an uncertainty of 0.71~ppm based on AGT method
 has been recently reported~\cite{dePodesta2013Metrologia_AGT_NPL}.
It also brings the concern that the new value of $k_B$ may be
 solely determined from AGT measurements.
In order to avoid the risk of unrevealed systematic deviation in single method,
 measurements using alternative methods other than AGT and with sufficiently low uncertainty ($<7$~ppm) is needed.

Doppler Broadening Thermometer (DBT) is an optical method to determine
 the product of $k_B$ and the thermodynamic temperature $T$
 from the Doppler width of a transition of atoms or molecules at thermodynamic equilibrium.
The Doppler width, $\Gamma_D$ (full width at half maximum, FWHM),
 relates with $k_B T$ following the equation:
\begin{equation} \label{EQ_DBT}
 \frac{\Gamma_D}{\nu_0} = \sqrt{8\ln 2\frac{k_B T}{m c^2}}
\end{equation}
In Eq.~\ref{EQ_DBT},
 $c=$ 299 792 458~m/s, the speed of light, is a constant without uncertainty,
 $m$ is the mass of the molecule,
 known with a relative accuracy of $10^{-8}$ for quite a few atoms and molecules,
 and the central frequency of the transition $\nu_0$ can be measured with a relative accuracy better than $10^{-9}$.
Therefore, precise measurements of the sample temperature $T$ and the Doppler width of the transition
 will result in a spectroscopic determination of $k_B$.
First DBT determination of $k_B$ was demonstrated by Daussy \textit{et al.} in 2007.
By measuring a NH$_3$ line near 10~$\mu$m, they determined the $k_B$ value with
 a relative uncertainty of $2\times$10$^{-4}$~\cite{Daussy2007PRL_DBT},
 and later on $5\times 10^{-5}$~\cite{Djerroud2009CRP_DBT, Lemarchand2011NJP_DBT}.
The Italian group at \textit{Seconda Universit\`{a} di Napoli}
 obtained an accuracy of 160~ppm using a CO$_2$ line at 2.006~$\mu$m~\cite{Casa2008PRL_DBT},
 and recently improved the accuracy to 24~ppm by measuring
 an absorption line of H$_2^{18}$O near 1.39~$\mu$m~\cite{Moretti2013PRL_DBT}.

The reported DBT results are mostly based on
 direct absorption spectroscopy of different atomic or molecular transitions,
 including Rb~\cite{Trung2011PRA_DBT_Rb},
 NH$_3$~\cite{Daussy2007PRL_DBT, Djerroud2009CRP_DBT, Lemarchand2011NJP_DBT},
 CO$_2$~\cite{Casa2008PRL_DBT},
 H$_2$O~\cite{Moretti2013PRL_DBT},
 C$_2$H$_2$~\cite{Yamada2009CRP_DBT, Hashemi2014JCP_DBT_C2H2},
 and O$_2$~\cite{Cygan2010PRA_DBT_O2}.
Molecular lines, as ro-vibrational transitions in most studies,
 have very narrow natural line width,
 and the saturation effect is also negligible.
However because they are usually much weaker than an atomic transition,
 higher sample pressures are required, and as a result,
 pressure-induced broadening should be taken into account in molecular DBT measurements.
In contrast, in a DBT measurement using atomic transitions,
 because atomic lines are usually stronger,
 which allows measurements at very low pressures,
 the pressure broadening is negligible,
  but the natural line width and power broadening must be considered.
In addition, transitions of closed-shell molecules are insensitive
 to electronic or magnetic field,
 while it must be carefully investigated in atomic studies.
We proposed to use very sensitive Cavity Ring-Down Spectroscopy (CRDS)
 for DBT measurements based on near-infrared molecular lines~\cite{Sun2011OE, Cheng2015CPB_DBT}.
CRDS was first implemented by
 O'Keefe and Deacon~\cite{Okeefe1988RSI_CRDS} in 1988.
It determines the absorption of gas samples
 by measuring the decay rate of the light emitted from
 a resonant cavity composed of two high-reflectivity (HR) mirrors.
Because photons travel between the HR mirrors many times
 before they escape from the cavity,
 the equivalent absorption path length is significantly enhanced.
The absorption coefficient $\alpha$ can be derived from the equation:
\begin{equation} \label{EQ_CRDS}
\alpha=\frac{1}{c}(\frac{1}{\tau}-\frac{1}{\tau_0})
\end{equation}
 where $\tau$ and $\tau_0$ are the ring-down time with and without absorption, respectively.

The ultra-high sensitivity of CRDS is particularly useful for DBT studies.
First, it allows to measure absorption spectra at a low pressure.
Since the spectrum should be recorded at certain pressures
 to acquire sufficient signal-to-noise ratio in DBT measurements,
 the line-shape in the recorded spectrum is a convolution of
  Doppler broadening and pressure broadening.
It is further complicated by speed-dependent collisions
 which correlates the Doppler shift and the collision-induced effects.
Despite that various line-shape models have been developed
 (see Refs.~\cite{Wcislo2013JQSRT_Profile, Cygan2013EPJST_Profile, Hartmann2013PRA_CO2_Profile} and references therein),
 it remains a great challenge to validate realistic line profiles
  from observed spectra.
Moretti \textit{et al.} reported the most precise DBT result
 to date~\cite{Moretti2013PRL_DBT}.
In their study, a leading uncertainty of 15~ppm is due to the line-shape model
 applied in fitting the spectra recorded at sample pressures of a few hundred Pa.
Therefore, it is necessary to record spectra at pressures as low as possible
 to minimize the influence from collision-induced effects.
Second, CRDS provides necessary high ``vertical'' resolution.
A DBT measurement with part-per-million (ppm) accuracy requires
 detecting spectral profiles with comparable precision.
CRDS has allowed an unprecedent sensitivity to the level of
 $10^{-11}$~cm$^{-1}$~Hz$^{-1/2}$.~\cite{Kassi2012JCP_CRDS, Truong2013NatPhoton_FARS_CRDS, Chen2014AO}
As a result, spectra with considerably high signal-to-noise ratio
 can be recorded by CRDS at very low sample pressures.
In CRDS, the detection of trace absorption is converted to monitoring
 changes in the decay time of the cavity (Eq.~\ref{EQ_CRDS}),
 which also leads to an enhancement of the dynamic range of the detection.
In addition, CRDS also allows to use a relatively smaller volume
 of sample gases compared to multi-pass configurations.
It eventually reduces the difficulty in maintaining the sample cell
 at an uniform temperature.

In this report, we present a CRDS instrument combined with a
 temperature-stabilized sample cavity devoted for DBT studies.
An infrared ro-vibrational transition of C$_2$H$_2$ was studied
 to demonstrate the capability of the instrument.
A statistical uncertainty of 6~ppm has been achieved in the
 line widths derived from the spectra recorded in a few hours.
The stability and reproducibility are investigated by
 measuring the spectroscopy determined temperatures in the range of 299 - 306~K.
The results indicate that a CRDS-based instrument is promising
 for DBT measurement toward the one-part-per-million accuracy.

\section{Experimental details}

\begin{figure}
    \includegraphics[width=3.5in]{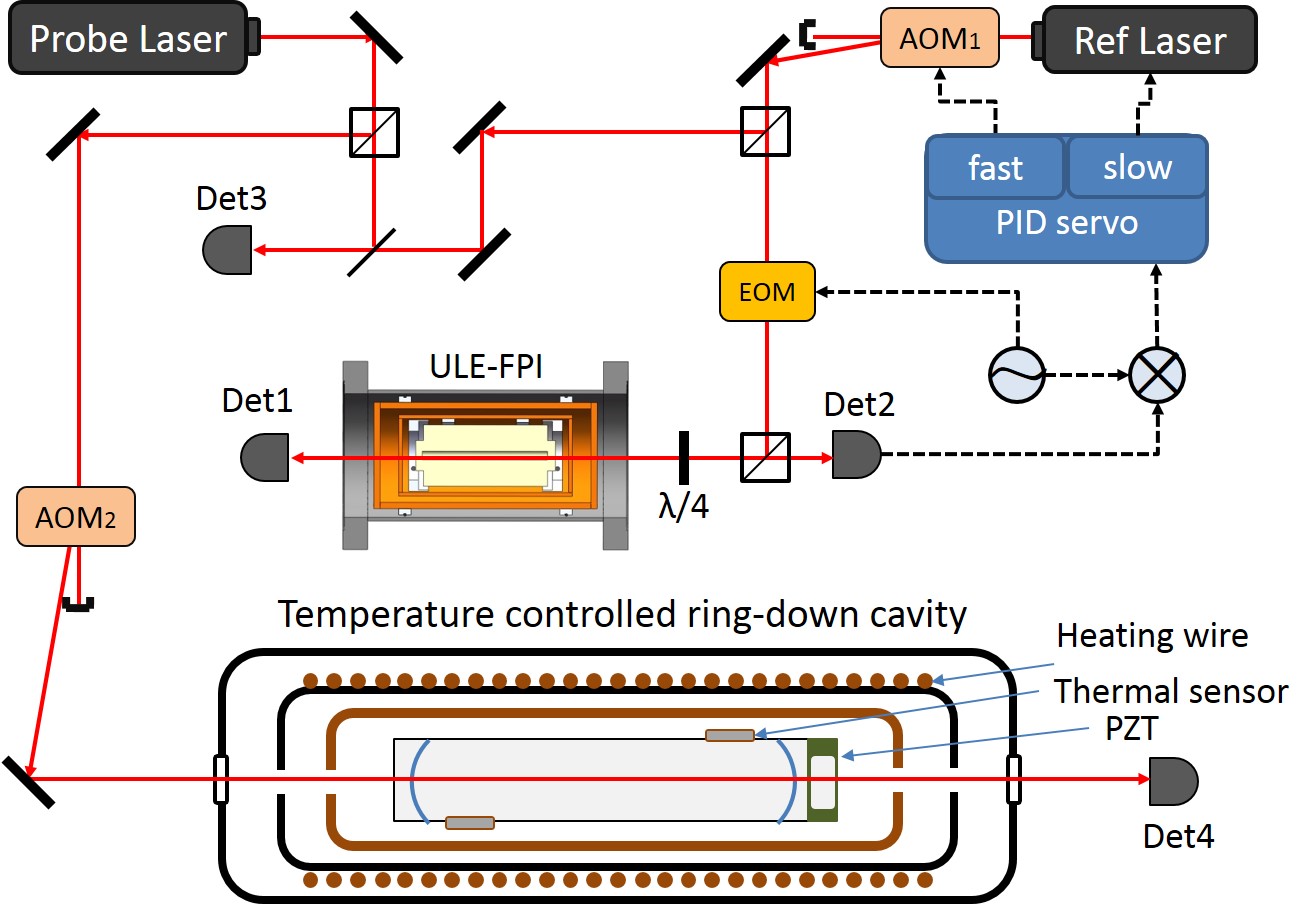}
    \caption{The configuration of the experimental setup for CRDS-DBT.
    The abbreviations are as following: AOM, acousto-optical modulator;
    EOM, electro-optical modulator;
    Det, detector;
    PZT, lead zirconate titanate piezoelectric actuator;
    ULE-FPI, Fabry-Perot interferometer made of ultra-low-expansion glass.
        \label{Fig_Setup}
    }
\end{figure}

The configuration of the experimental setup is presented
 in Fig.~\ref{Fig_Setup}.
The spectroscopy part of the instrument is close to that given in
 Refs.~\cite{Pan2011RSI_LL_CRDS, Cheng2012OE}.
A reference laser is locked on a longitudinal mode of a Fabry-P\'{e}rot interferometer (FPI)
 using the Pound-Drever-Hall method~\cite{Drever1983APB_PDH}.
The slow and fast feed-back control signals are delivered to
 the laser controller and an acousto-optical modulator (AOM1), respectively.
The FPI is made of ultra-low-expansion (ULE) glass,
 installed in a vacuum chamber,
 and thermo-stabilized at $29^\circ$C,
 a magic temperature that the thermo-expansion coefficient of the ULE-FPI is close to zero.
The frequency drift of the longitudinal modes of the ULE-FPI
 has been estimated to be less than 1~kHz
 by comparing to atomic transitions~\cite{Cheng2012OE}.
A Ti:Sapphire laser (Coherent MBR 110) is used as the probe laser
 for cavity ring-down spectroscopy.
The line width of the probe laser is about 75~kHz stated by the manufacturer.
The beat signal between the probe laser and the reference laser
 is locked to a RF synthesizer (Agilent N9310A)
 referenced to the GPS signal (Spectratime GPS Reference-2000).
Another AOM (AOM2) is used as an optical switch.
The total laser power sent to the ring-down cavity is about 10~mW.
The high-finesse ring-down cavity is composed of
 a pair of mirrors with a reflectivity of 0.99995.
The light emitted from the cavity is detected by an avalanche photo-detector.
The bandwidth of the detection line is 15~MHz.
Once the detected signal reaches a preset threshold,
 a trigger signal will be produced,
  to shut off the input laser beam using AOM2
  and also to start recording the ring-down event with an AD converter installed in a computer.

\begin{figure}
    \includegraphics[width=3.5in]{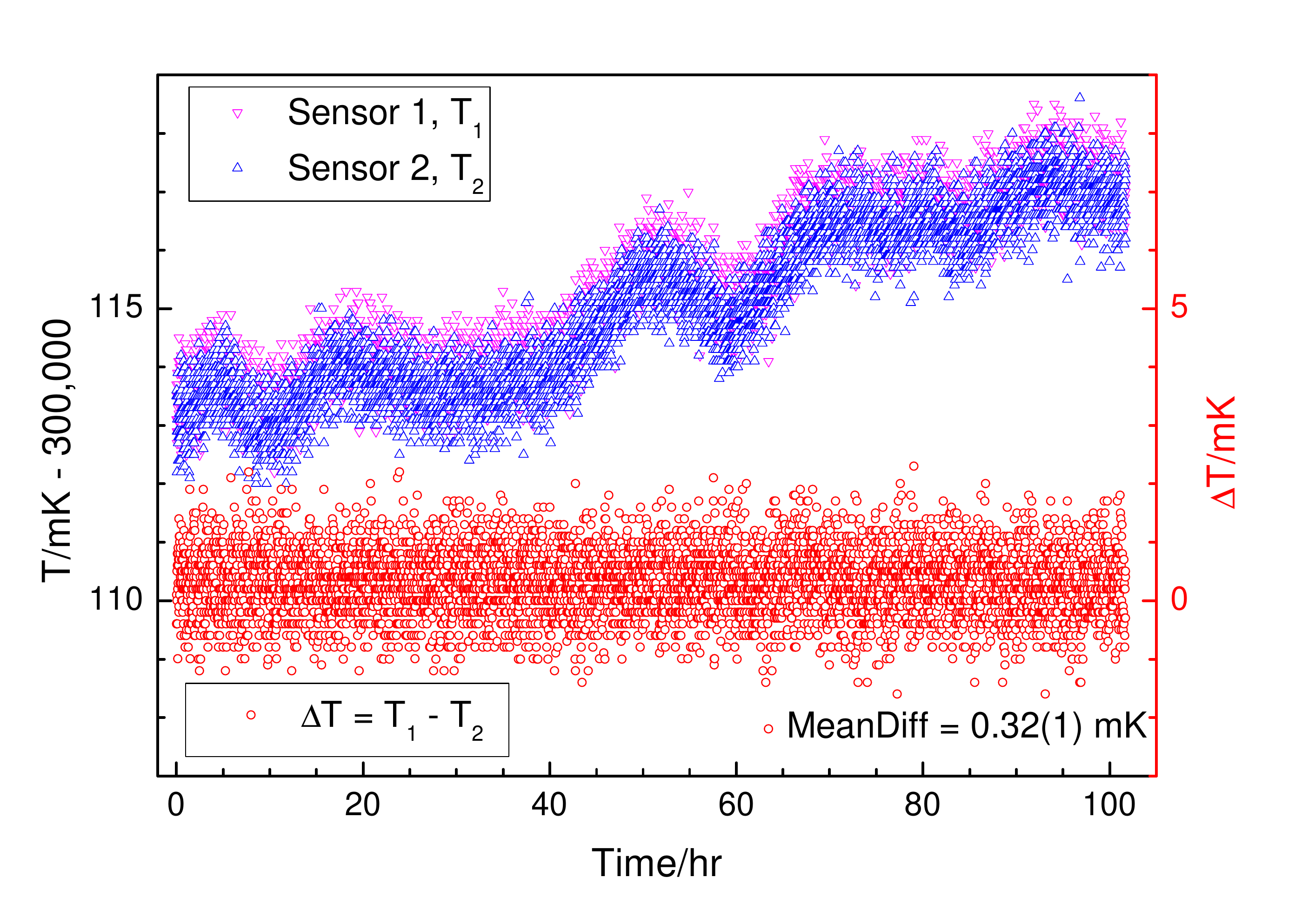}
    \caption{ Temperature of the ring-down cavity recorded by two thermal sensors during 100 hours.
    These two sensors are attached at both ends of the cavity, separated by 40~cm.
    The temperatures obtained from two sensors are shown with triangles
     and their differences are shown with circles.
        \label{Fig_Temp}
    }
\end{figure}

The ring-down cavity, 50~cm long, made of aluminum,
 is installed in a vacuum chamber.
Two layers of aluminum shields are used between the ring-down cavity
 and the vacuum chamber.
On the outside layer, two heating wires respectively
 controlled by two feedback circuits
 are used to maintain a temperature stability of about 10~mK.
The inner layer is used as a heat shield which also helps
 to reduce the temperature gradient along the cavity.
Two platinum thermal sensors are attached on the wall
 at both ends of the 50-cm-long ring-down cavity, separated by 40~cm.
The sensors and the readout (MKT50, Anton Parr) have been
 calibrated in National Institute of Metrology (Beijing, China).
Fig.~\ref{Fig_Temp} shows the recorded temperatures given
 by the two sensors.
The temperature fluctuation was less than 5~mK during 100 hours.
The recorded temperature difference between two sensors
 is only 0.3~mK, which is actually below the calibration accuracy (0.5~mK).
It indicates an uniform temperature along the whole ring-down cavity.

Acetylene gas sample was bought from Nanjing Special Gas Co.,
 with a stated purity of 99.5\%.
The sample was purified by the ``freeze-pump-thaw'' method before use.
The sample pressure is measured by a capacitance manometer (MKS 627B) with
 a full-scale range of 133~Pa.
Because a sample pressure of only 1-2 Pa was used in this study,
 instead of the reading from the manometer,
 the partial pressure of C$_2$H$_2$ was determined from the integrated absorption line intensity of C$_2$H$_2$ and the line strength values reported by
 Herregodts \textit{et al.}~\cite{Herregodts1999JCP_C2H2, Herregodts2003MP_C2H2}
The R(9) line in the $\nu_1+3\nu_3$ band of $^{12}$C$_2$H$_2$
 is selected as the ``target'' line for DBT measurement.
The transition frequency has been precisely determined to be
 12696.412751(16)~cm$^{-1}$, with a relative precision of
 $1.3\times 10^{-9}$.~\cite{Liu2013JCP_C2H2}

\section{Results and discussion}

\begin{figure}
    \includegraphics[width=3.5in]{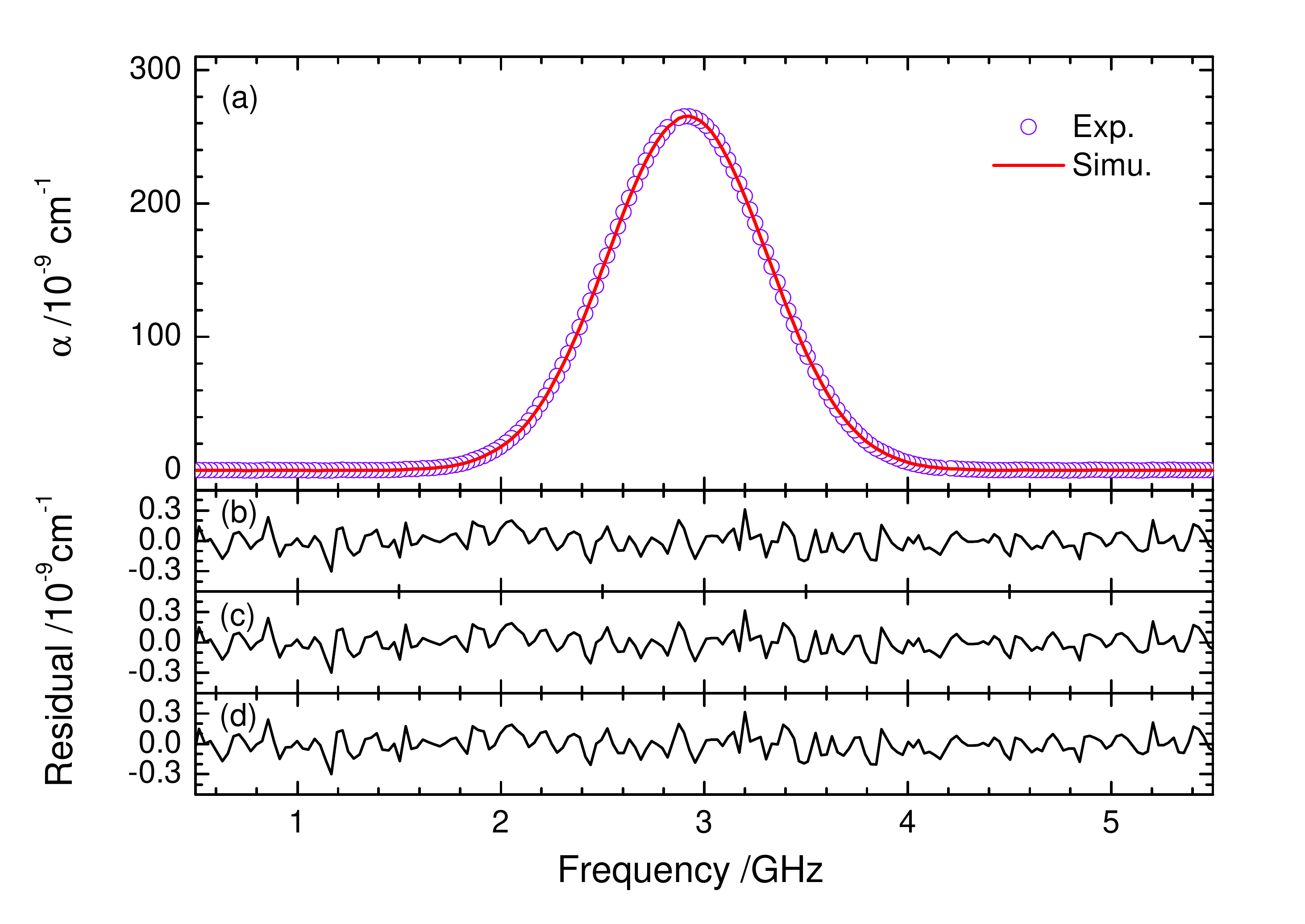}
    \caption{The $R(9)$ line in the $\nu_1$+$3\nu_3$ band of $^{12}$C$_2$H$_2$.
    The sample vapor pressure is 1.54~Pa.
    (a) The experimental data (open circles) and the simulated spectrum (solid curve).
    The residuals from fitting using different profiles are given:
    (b) Gaussian,
    (c) Voigt, with the Lorentzian width (FWHM) fixed at 0.709~MHz,
    (d) Rautian, with the Lorentzian width fixed at 0.709~MHz
     and the Dicke narrowing coefficient fixed at 0.018~MHz.
        \label{Fig_Spec}
    }
\end{figure}

\subsection*{Statistical uncertainty and reproducibility}

An example of the recorded spectrum of the C$_2$H$_2$ line
 at 787.6~nm is shown in Fig.~\ref{Fig_Spec}.
The sample pressure was 1.54~Pa derived from the measured integrated line intensity.
It took about 2 minutes to record the spectrum (one scan).

\begin{figure}
    \includegraphics[width=3.5in]{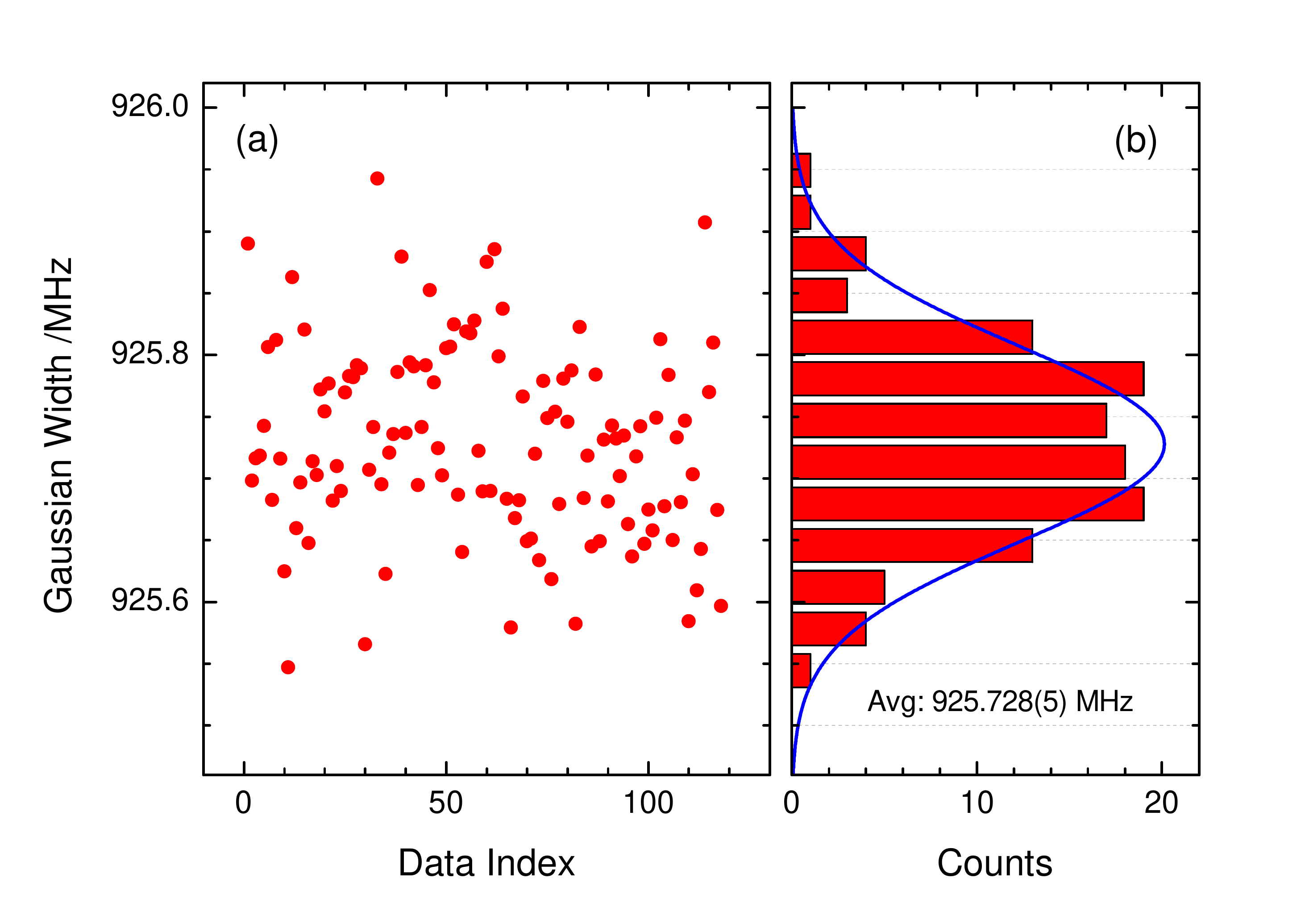}
    \caption{
    (a) Gassian width derived from each C$_2$H$_2$ spectrum.
    A Voigt profile with the Lorentzian width fixed at 0.709~MHz
     was applied in the fitting.
    (b) Statistics of the obtained Gaussian width $\Gamma$.
    A Gaussian fit of the counts of the $\Gamma$ values
     gives an averaged value of $925.728\pm 0.005$~MHz.
    \label{Fig_Widths}
    }
\end{figure}

The line profile should be a composite of Doppler broadening, pressure broadening,
 transit-time broadening, and power broadening,
 denoted as (FWHM) $\Gamma_D$, $\Gamma_P$, $\Gamma_T$ and $\Gamma_S$, respectively.
In our measurement, the radius of the laser beam in the ring-down cavity is 0.3~mm,
 corresponding to a transit-time width of 0.365~MHz at 300~K.
Ma \textit{et al.}~\cite{Ma1999JOSAB_C2H2} reported
 a pressure broadening coefficient of $29.8\pm 0.1$~MHz/Torr
 (0.344~MHz at 1.54~Pa)
 from analysis of saturation spectroscopy carried out at milli-Torr pressures.
Ma \textit{et al.} also reported that the saturation power of
 the C$_2$H$_2$ lines near 790~nm is a few Watts.
In this study, according to the amplitude of the signal detected by the detector
 (Det4 in Fig.~\ref{Fig_Setup}),
 we estimate the emitted laser power from the cavity is about 1~$\mu$W.
Taking into account the enhancement of the resonance cavity ($F\sim 60000$),
 the light power built up in the ring-down cavity is about 60~mW,
 far from the power needed to saturate the transition.
Therefore the power broadening is neglected in this study.
It is also worth noting that ``heating'' of the gas due to absorption is negligible
 because the transition is very weak ($\alpha < 10^{-6}$/cm).
Consequently, the Lorentzian width is 0.709~MHz, a sum of $\Gamma_T$ and $\Gamma_P$.

We fit the spectrum using different profiles, Gaussian, Voigt, and Rautian,
 and the fitting residuals are shown in Fig.~\ref{Fig_Spec}.
Note that the amplitudes of the fitting residuals are similar among the fits using
 different line profile models.
A Gaussian width $\Gamma_D$ can be derived from the fit of each spectrum.
About 120 values of $\Gamma_D$ derived from
 fitting the spectra recorded in several hours are depicted in Fig.~\ref{Fig_Widths}.
Voigt profile with Lorentzian width fixed at 0.709~MHz was applied in the fitting.
A statistics of the values gives an averaged Gaussian width of
 $925.728\pm 0.005$~MHz, with a relative uncertainty of 6~ppm.

In order to investigate possible systematic deviations in the results,
 we also treat the same data using different line shape models.
The results from fitting with pure Gaussian and Rautian profiles
 are given in Table~\ref{Table_Temp}.
As expected, the pure Guassian profile tends to overestimate the Doppler width.
Note that Herregodts \textit{et al.} reported a pressure self-broadening coefficient
 of about 0.9~GHz/atm (FWHM) according to their measurement
 at sample pressures of a few hundred Torr.~\cite{Herregodts2003MP_C2H2}
It corresponds to a Lorentzian width of 0.137~MHz at 1.54~Pa,
 which is much less than the value we applied in the analysis above.
Since the sample pressure used in this study is at the milli-Torr level,
 close to that used in  Ref.~\cite{Ma1999JOSAB_C2H2},
 we chose to use the coefficient given in Ref.~\cite{Ma1999JOSAB_C2H2}.
As a comparison, we tried to fit the spectra again
 using a $\Gamma_P$ value of 0.137~MHz,
 and the results are also given in Table~\ref{Table_Temp}.
The differences among the results from different fitting conditions
 could be used to estimate the deviation from different sources.

The transit-time broadening has a major contribution in the uncertainty.
In our measurement, the radius of laser beam in the ring-down cavity is
 calculated according to the optical configurations,
 which may have a relative deviation as high as 10~\%.
According to our numerical analysis, it will induce an uncertainty of 10~ppm or less
 in the derived $\Gamma_D$ value,
 which can also be easily estimated from the values given in Table~\ref{Table_Temp}.
The pressure-broadening coefficient reported in Ref.~\cite{Ma1999JOSAB_C2H2}
 has a relative uncertainty of about 0.3\%,
 and the sample pressure determined from the absolute line intensity~\cite{Herregodts2003MP_C2H2} has an uncertainty of less than 4\%,
 therefore the resulted uncertainty in $\Gamma_P$ is about 0.013~MHz,
 which leads to a relative uncertainty of about 7~ppm in $\Gamma_D$.
However, as shown in Table~\ref{Table_Temp}, if the pressure-broadening coefficient
 derived from high-pressure measurements~\cite{Herregodts1999JCP_C2H2, Herregodts2003MP_C2H2} is applied,
 the change in resulted $\Gamma_D$ value is as high as 120~ppm.
It indicates that the pressure-broadening coefficient
 need to be carefully investigated in future DBT measurement.

The difference between the Voigt and Rautian profiles,
 about 20~ppm in derived $\Gamma_D$ values shown in Table~~\ref{Table_Temp},
 can be a rough estimation of the uncertainty rising from the profile model.
As have been intensively studied in
 Refs.~\cite{Moretti2013PRL_DBT, Borde2009CRP_DBT_Profile, Cygan2010PRA_Lineshape_Boltzmann, Gianfrani2012JPCS_DBT_Profile, Triki2012PRA_DBT_Profile_NH3, Rohart2014PRA_DBT_Profile},
 the uncertainty can be significantly reduced by using more sophisticated line profile models.
Moreover, by measuring the spectra at different pressures and extrapolate
 the results to the zero-pressure limit,
 it is possible to reduce the systematic error of the Doppler width to less than 1~ppm.~\cite{Cygan2010PRA_Lineshape_Boltzmann}
Moretti \textit{et al.} have estimated that the uncertainty due to line profile models
 is about 15~ppm for their measurements with sample pressures at the $10^2$~Pa level.~\cite{Moretti2013PRL_DBT}
Since the sample pressure used in our CRDS measurement is about 1/100 of that used in
 conventional direct absorption studies,
 we expect that the uncertainty could be potentially reduced to less than 1~ppm.

\begin{table}
\caption{Gaussian widths (FWHM, in MHz) and corresponding
 spectroscopy temperatures (in K) derived from the fitting of the spectra
 using different line profile models and parameters.
 The Lorentzian width ($\Gamma_T+\Gamma_P$) is fixed in the fitting.
 \label{Table_Temp}
}
\begin{tabular}{ccccc}
\hline%
 Model & $\Gamma_D$ (MHz) & $\Gamma_T$ & $\Gamma_P$ & $T_{spec}$ (K)\\
\hline%
Gaussian& 926.132(6) & -     & -     & 300.240(4)\\
Voigt   & 925.728(5) & 0.365 & 0.344 & 299.978(3)\\
Rautian & 925.744(5) & 0.365 & 0.344 & 299.988(3)\\
Voigt   & 925.844(6) & 0.365 & 0.137 & 300.053(4)\\
Rautian & 925.861(6) & 0.365 & 0.137 & 300.064(4)\\
\hline%
\end{tabular}

\end{table}

\begin{figure}
    \includegraphics[width=3.5in]{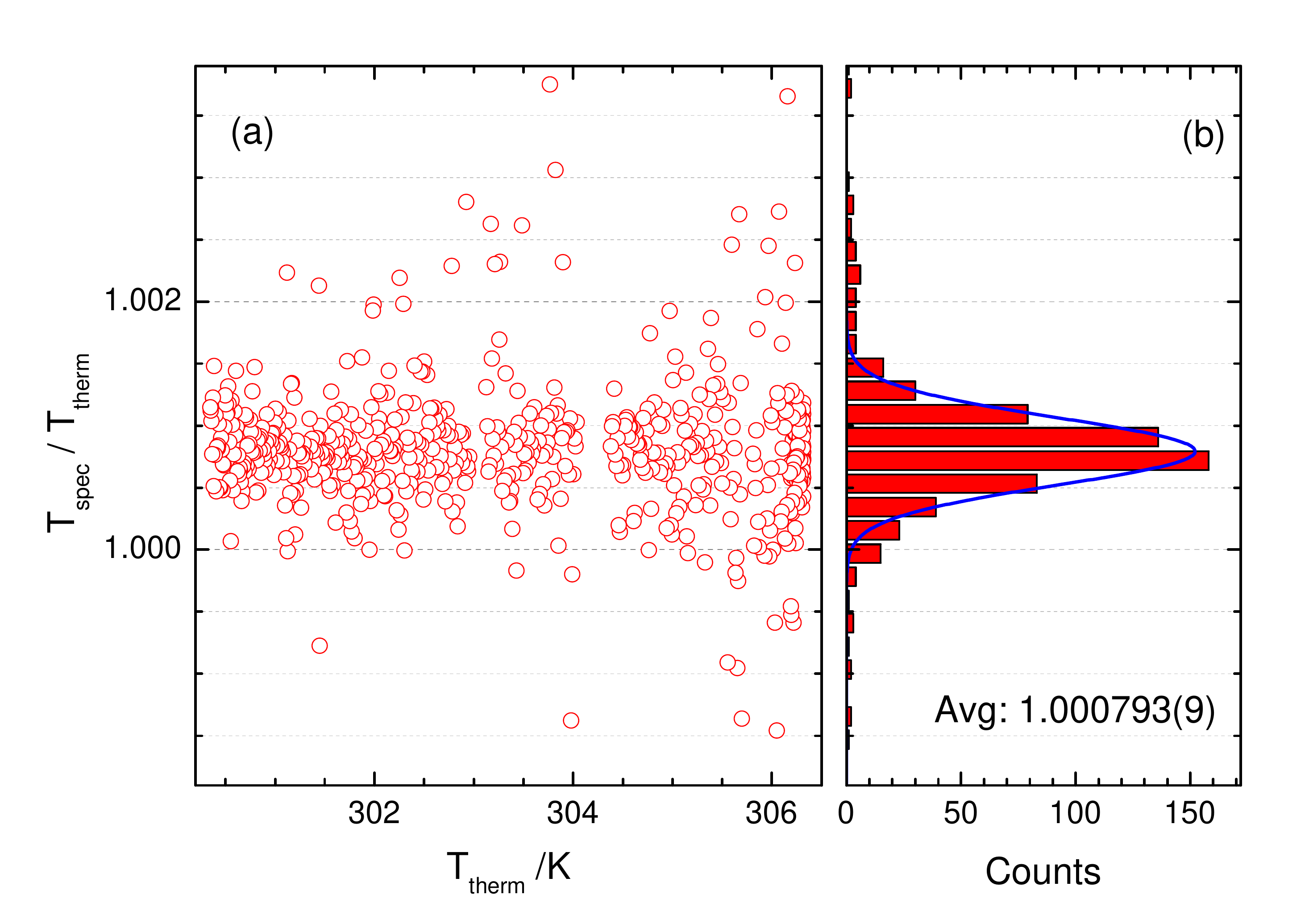}
    \caption{
    (a) Comparison of the temperatures derived from the C$_2$H$_2$ spectra ($T_{spec}$) and that from the thermal sensors ($T_{therm}$).
    (b) Statistics of the $T_{spec}/T_{therm}$ values.
    \label{Fig_dT}
    }
\end{figure}

The stability and reproducibility of the instrument are investigated
 by more measurements carried out at different temperatures
 between 299~K and 306~K.
Both temperatures derived from the C$_2$H$_2$ spectra ($T_{spec}$)
 and that from the thermal sensors ($T_{therm}$),
 are shown in Fig.~\ref{Fig_dT}.
A statistics of the 626 data in total shows that
 the mean ratio of $T_{spec}/T_{therm}$ is 1.000~793(9).
The relative statistical uncertainty is 9~ppm,
 agreeing with the uncertainty from the spectroscopy measurement.
However, there is a systematic deviation of 793~ppm,
 which is considerably larger than expected:
 the uncertainty in $T_{therm}$ should be less than 10~ppm,
 and we have shown that the deviation due to
 improper line profiles or spectroscopic parameters should be less than 200~ppm.

\subsection*{Interference due to ``hidden'' lines}

\begin{figure}
    \includegraphics[width=3in]{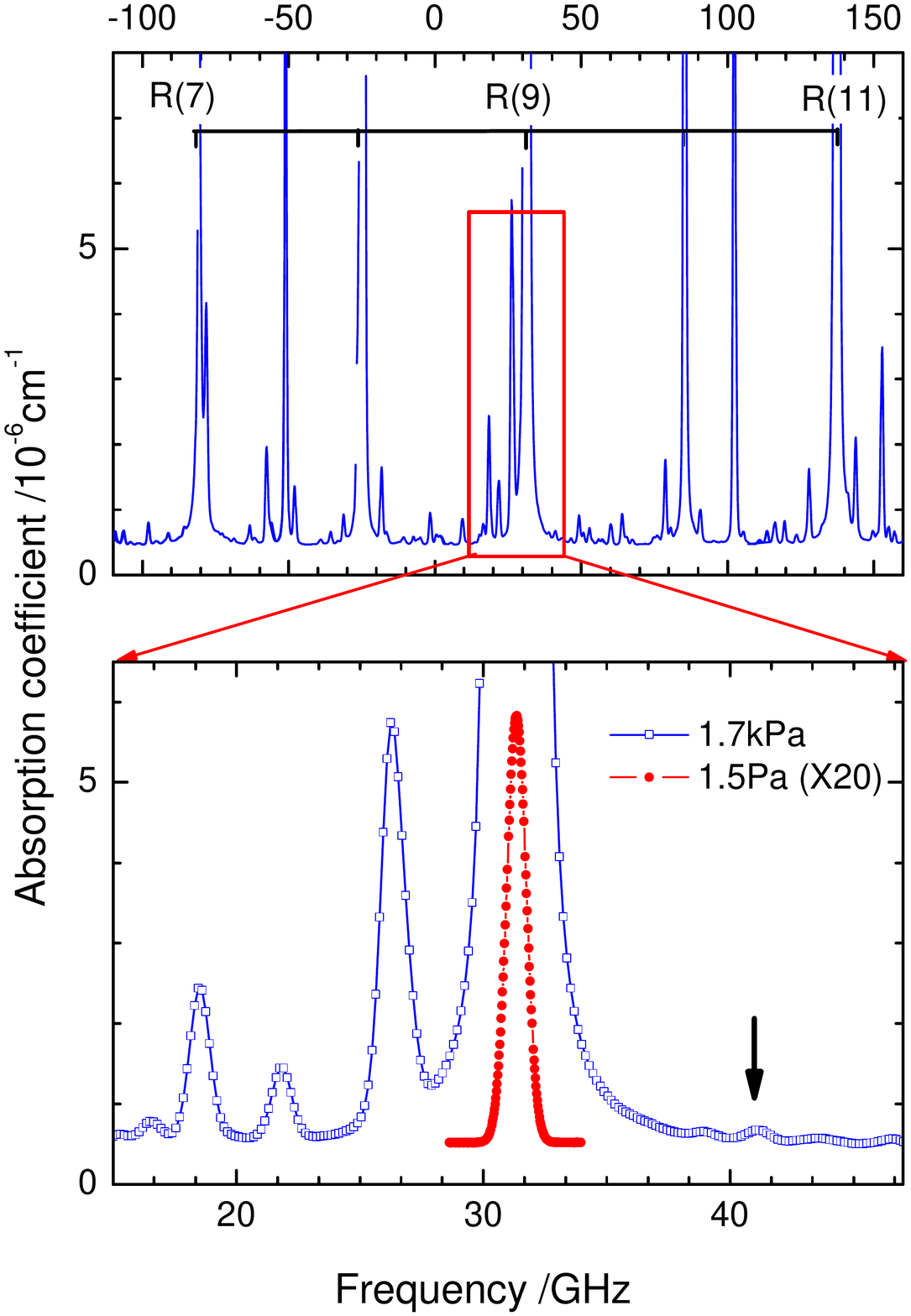}
    \caption{Cavity ring-down spectrum of C$_2$H$_2$ near 787~nm.
    The R(7) - R(11) lines in the $\nu_1+3\nu_3$ band are marked.
    Note the central parts of these lines are out of the dynamic range
     at a sample pressure of 1.7~kPa.
    The lower panel shows the spectrum close to the R(9) line.
    For comparison, the spectrum recorded at 1.54~Pa is also shown
     in the lower panel (has been multiplied by a factor of 20).
    \label{Fig_HighP}
    }
\end{figure}

The reason could be that the spectrum shown in Fig.~\ref{Fig_Spec}
 is not really due to an isolated C$_2$H$_2$ line.
We carried out a spectral scan around this line
 with much higher acetylene sample pressures.
A piece of the spectrum recorded at 1.7~kPa is shown in Fig.~\ref{Fig_HighP}.
At such a high pressure, the central parts of the strong lines
 shown in the figure are out of the dynamic range.
The positions of these strong lines are marked in Fig.~\ref{Fig_HighP},
 and they have been reported in our previous study~\cite{Liu2013JCP_C2H2}.
As shown in the figure,
 many weak C$_2$H$_2$ lines are located in the region.
They may be lines of different vibrational bands,
 hot bands, or due to minor isotopologues,
 and their positions have not been reported before.
Due to the high density of lines, there could be very likely weak lines
 ``hidden'' in the vicinity of the much stronger R(9) line.
The situation is similar for other strong lines in the $\nu_1+3\nu_3$ band.
The interference due to a ``hidden'' weak line overlapped with the target line
 but not included in the spectral fitting
 will lead to considerably overestimated Gaussian line width from the fitting.
Note that this deviation cannot be removed by accumulating more measurements
 in different pressures.

\begin{figure}
    \includegraphics[width=3in]{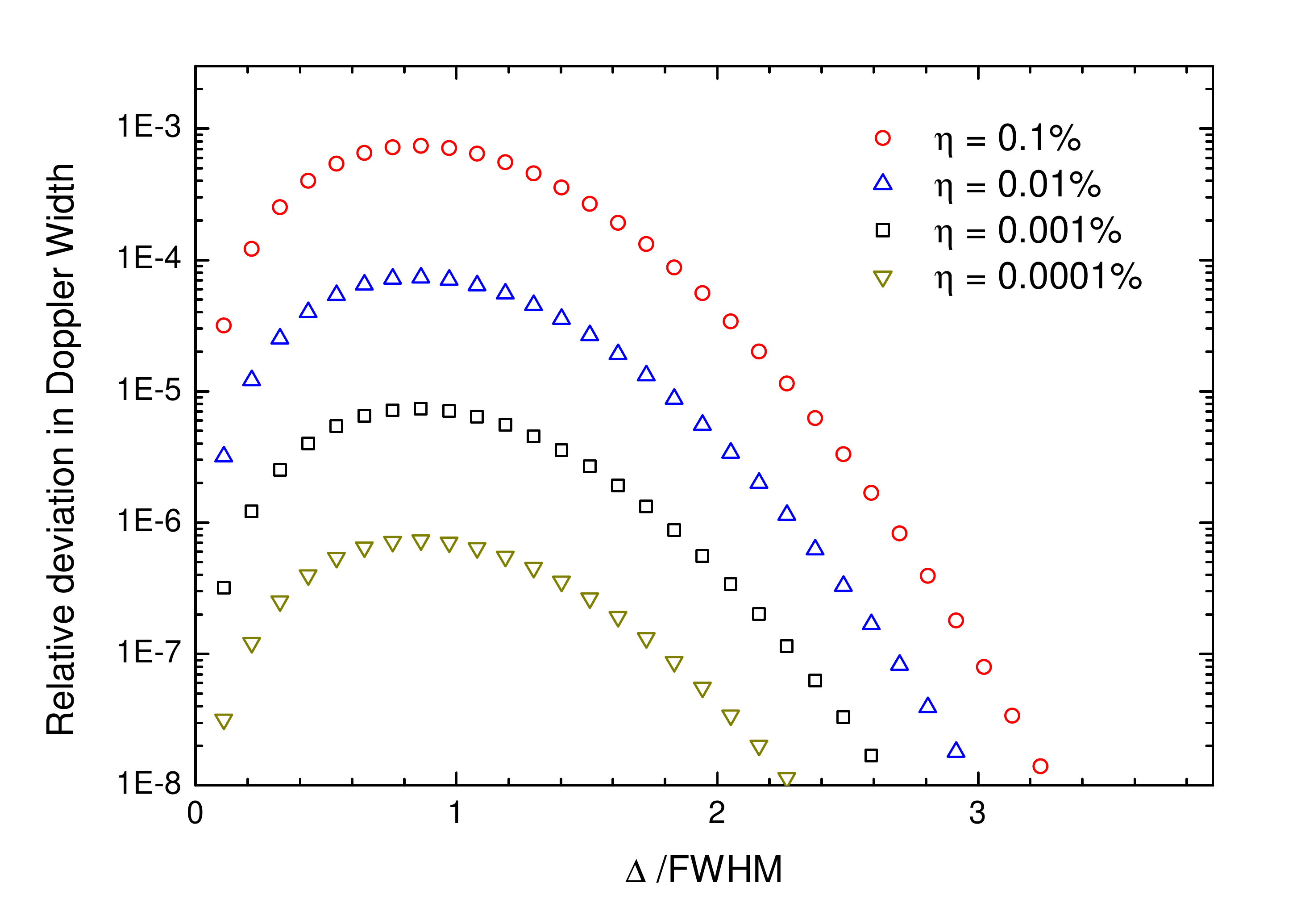}
    \caption{Influence on the derived Doppler width
    from a ``hidden'' weak line close to the strong ``target'' line.
    $\Delta$ is the distance between these two lines,
    and $\eta$ is the relative strength of the weak line.
    The strength of the strong line is normalized as 1.
    \label{Fig_SimuHidden}
    }
\end{figure}

In order to give a quantitative inspection of the influence from the ``hidden'' lines,
 we produced a series of simulated spectra with a weak ``hidden'' line
 close to the strong ``target'' line.
The weak line has a relative strength of $\eta$,
 and a distance of $\Delta$ from the strong line.
For simplicity, both lines are only Doppler broadened.
Under different $\eta$ and $\Delta$ values,
 the simulated spectrum is fitted with a single Gaussian peak and
 the deviation of the derived Gaussian width from the true value is shown in Fig~\ref{Fig_SimuHidden}.
As shown in the figure, within the distance of about three times of the Doppler width $\Gamma_D$,
 the presence of the weak line could significantly distort the derived Doppler width.
For example, a ``hidden'' line with a $\eta=0.1$\%, which has a strength close to
 that of weak line indicated with an arrow on Fig.~\ref{Fig_HighP},
 could lead to a relative deviation in $\Gamma_D$ of several hundred ppm
 if the line is close enough (within $3\Gamma_D$) to the target line.
From the spectrum of C$_2$H$_2$ shown in Fig.~\ref{Fig_HighP},
 due to the limited knowledge of the complicated spectrum in this region,
 we can hardly rule out the possibility of the existence of the unknown weak lines
 close to the ``target'' R(9) line.
Since the noise level shown in Fig.~\ref{Fig_Spec} is about 0.1\%,
 it indicates such ``hidden'' lines may also be at the level of $\eta\sim0.1$\%.
Therefore, we estimate the deviation due to this effect to be about 1000~ppm.

\begin{figure}
    \includegraphics[width=3.5in]{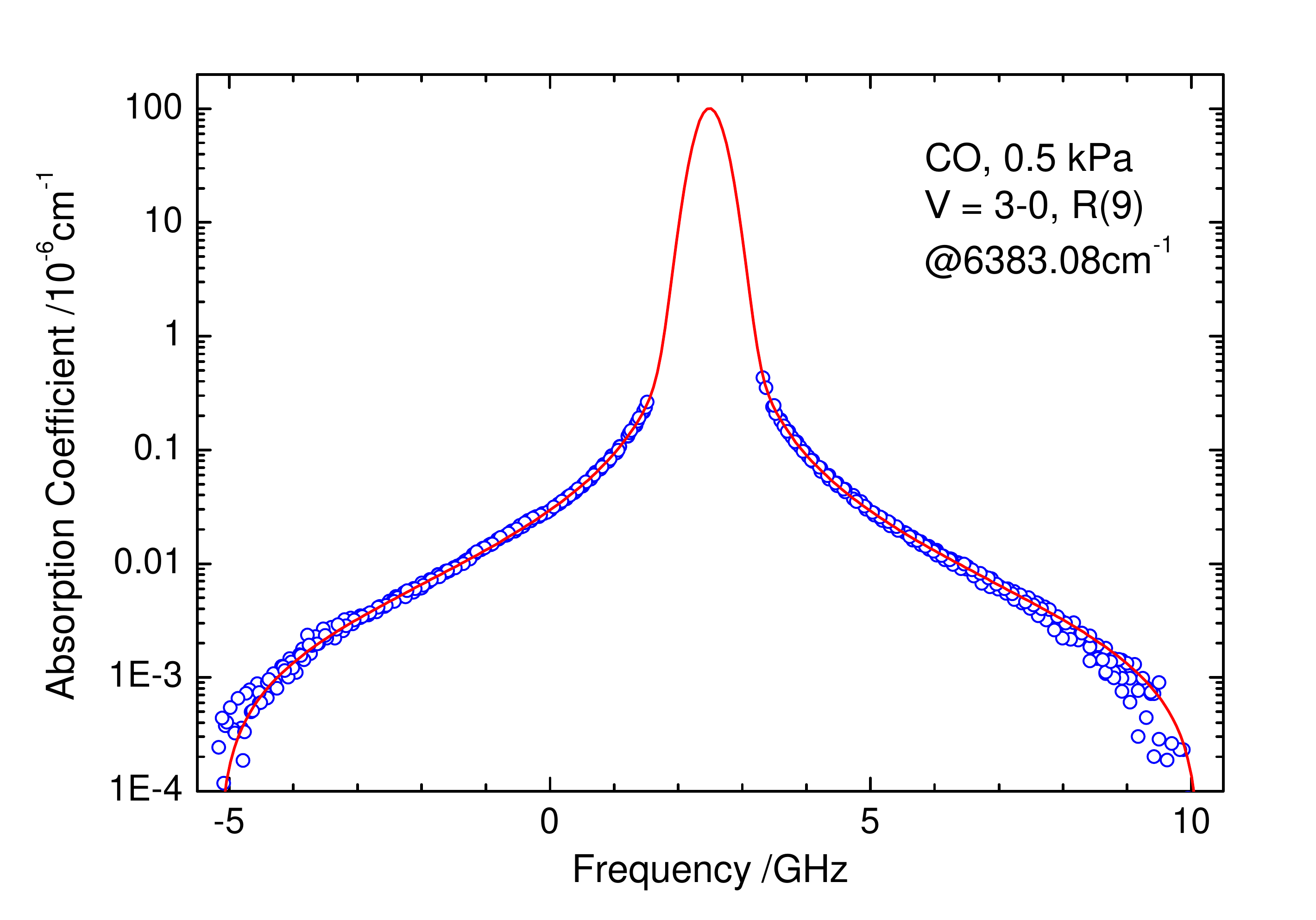}
    \caption{Cavity ring-down spectrum of CO near 1567~nm.
    Circles are experimental data, and solid line is for a simulated spectrum.
    The central part of the R(9) line in the $V=3-0$ band is out of the dynamic range.
    \label{Fig_CO}
    }
\end{figure}

In this respect, it is necessary to use a really isolated line
 for precise DBT measurements.
For a polyatomic molecule like C$_2$H$_2$,
 the density of ro-vibrational transitions is very high,
 even for lower vibrational states
 (see example in a recent study~\cite{Lyulin2014MP_C2H2_CRDS}).
Insufficient knowledge on the weak transitions will make it difficult to
 secure a precision of 1~ppm in DBT determination of $k_B$,
 since one needs to consider all the nearby lines with strengths
 in a dynamic range of six orders of magnitudes.
We plan to replace the ``target'' line with a ro-vibrational line of CO
 in the 1.5~$\mu$m region.
There are several advantages to use the CO molecule.
First, the ro-vibrational transitions of CO have been well studied,
 including very weak hot bands and the lines from minor isotopologues~\cite{Li2015APJSS_CO, Mondelain2015JQSRT_CO_V3},
 which allows to use a truly isolated line for high-precision DBT studies.
Second, the carbon monoxide gas sample can be easily purified
 in laboratory by removing the contaminant gases with a cold trap.
In contrast, other carbon hydrides are often detectable in acetylene samples,
 which will further complicates the spectrum.
Fig.~\ref{Fig_CO} shows the spectrum of the R(9) line in the second overtone of $^{12}$C$^{16}$O recorded by a CRDS instrument based on a
 distributed feed-back diode laser~\cite{Chen2014AO}.
The line is located at 6383.08~cm$^{-1}$, with a line intensity of $2.034\times 10^{-23}$~cm/molecule,
 which is similar to the C$_2$H$_2$ line shown in Fig.~\ref{Fig_Spec}.
Because the central part of the line is too strong to be detected
 at a sample pressure of 0.5~kPa,
 only wings of the line are shown in Fig.~\ref{Fig_CO}.
By comparing the observed spectrum and a simulated one,
 we cannot find any evidence of other unknown transitions of CO
 within a range of 20~GHz around the line center.
Therefore, this line could be a good candidate for DBT measurements.

\subsection*{Uncertainty budget}

A summary of the uncertainty budget is given in Table~\ref{Table_Budget}.
They are discussed as follows. \\
(1) The present statistical experimental uncertainty in the determination of the
 Doppler width is 6~ppm, which will leads to a relative uncertainty of 12~ppm
  in determined $k_B$ (note that $\delta k_B \approx 2\delta\Gamma_D$).
 The line strength of CO at 1.57~$\mu$m is very close to
  the C$_2$H$_2$ line at 787~nm used here,
  and we can extend the measurement time from several hours to several hundreds hours,
  we expect to reduce the uncertainty to 4~ppm or less.
 It has been proved to be effective that accumulating scans can lead to
  a decreasing noise level in CRDS measurements~\cite{Kassi2012JCP_CRDS, Chen2014AO}. \\
(2) The optical frequency is calibrated by the longitudinal modes of an ultra-stable
 Fabry-P\'{e}rot interferometer and microwave source,
 both with stability better than 1~kHz,
 therefore the uncertainty from the frequency calibration is negligible.\\
(3) The temperature measurement has an uncertainty of about 3~mK at room temperature.
 The uncertainty can be reduced to less than 0.3~mK when the measurement is
  carried out at the triple point of water.
 A sample cell stabilized at the temperature of the triple point of water is under test,
  and the temperature uncertainty is expected to be less than 1~ppm (0.3~mK). \\
(4) The relative accuracy of the center frequency of the C$_2$H$_2$ line
 is about 1.3~ppb ($\delta\nu_0/\nu_0$).
 For the CO lines at the 1.57~$\mu$m which will be used in succeeding studies,
  the line centers have been determined to sub-MHz accuracy~\cite{Mondelain2015JQSRT_CO_V3},
  therefore the induced uncertainty on $k_B$ is negligible. \\
(5) Recently Bord\'{e} has shown that the transit-time broadening is absent in
 linear absorption spectroscopy in the case of a uniform and isotropic medium~\cite{Borde2009CRP_DBT_Profile}.
In our measurements, the light power is at the level of a few percents of the saturation power and
 the contribution from the transit-time effects needs to be further investigated.
In present study, when we take into account the transit-time broadening,
 the uncertainty in $k_B$ induced by the uncertainty in $\Gamma_T$ is about 20~ppm.
It is mainly due to the 10~\% uncertainty in the radius of the laser beam in the cavity.
By a careful analysis of the optical configuration and
 extrapolating the results obtained at different configurations
 to the limit of zero transit-time broadening,
 we expect that the uncertainty could be reduced to 2~ppm or less. \\
(6) As discussed above, the current systematic uncertainty due to
 using improper line profile model is about 40~ppm in this work.
 By using more sophisticated line profiles and extrapolating the results obtained
  at different pressures to the zero-pressure limit~\cite{Moretti2013PRL_DBT, Borde2009CRP_DBT_Profile, Cygan2010PRA_Lineshape_Boltzmann, Gianfrani2012JPCS_DBT_Profile, Triki2012PRA_DBT_Profile_NH3, Rohart2014PRA_DBT_Profile},
  the uncertainty could be reduced to 1~ppm or less. \\
(7) The dominant contribution to the uncertainty in present study is
from the interference of ``hidden'' weak lines.
 As discussed above, by using a really isolated CO line at 1.57~$\mu$m,
  this influence can be dramatically reduced to 1~ppm or less. \\
(8) The laser power built in the cavity is estimated as about 60~mW,
 which is only about 1/100 of the saturation power needed for the C$_2$H$_2$ transition.
 Therefore the contribution from the saturation effect is negligible. \\
(9) The line width of the probe laser source also contributes to the uncertainty.
 The Ti:saphire laser used in this study has a line width of 75~kHz,
  which is about $8\times 10^{-5}$ of the Doppler width.
 As we have shown in previous analysis~\cite{Sun2011OE}, it may cause an uncertainty of about 10~ppm
 in $k_B$.
 A narrow-band fiber laser with a line width of 0.1~kHz will be used
  in the measurement of the CO line at 1.57~$\mu$m,
  and we can expect that this effect would also be eliminated. \\
(10) $^{12}$C$^{16}$O has no hyperfine structure. \\
(11) As has been discussed in our previous study~\cite{Sun2011OE},
 the uncertainty due to detector nonlinearity can be reduced
  to less than 1~ppm in CRDS-based DBT measurement.

In total, the uncertainty in determined $k_B$ value is 12~ppm (type A)
 and about 1000~ppm (type B) in current study.
We expect an uncertainty reduced to 4~ppm or less with an upgraded system
 using a CO transition at 1.57~$\mu$m as the ``target'' line.

\begin{table}
\caption{Uncertainty budget in the CRDS determination of $k_B$ (ppm).
 \label{Table_Budget}
}
\begin{tabular}{lcc}
\hline%
 Contribution & Current & Upgrade \\
\hline%
(1) Experimental statistical  & 12* &$<4$\\
(2) Frequency calibration     &$<1$& negligible \\
(3) Sample temperature        &10  &$<1$\\
(4) Line center frequency     & &negligible \\
(5) Transit-time broadening   &20  &$<2$  \\
(6) Line shape model          &40  & 1  \\
(7) ``hidden'' line           &$\sim 1000$ & $<1$\\
(8) Saturation broadening     & &negligible  \\
(9) Laser linewidth           &10   &$<0.1$\\
(10) Hyperfine structure       & & negligible \\
(11) Detector nonlinearity     &$<1$ & negligible \\
Total uncertainty         &1000 &$<5$\\
\hline%
\end{tabular}
\flushleft
* Type A uncertainty, others are Type B uncertainties.
\end{table}

\section{Conclusion}

Cavity ring-down spectroscopy (CRDS) can be applied as an optical thermometry
 by measuring the Doppler width of an absorption line of atoms or molecules.
Its high sensitivity allows to detect precise line profiles
 at relatively low sample pressures.
As a demonstration, using a thermo-stabilized ring-down cavity,
 the acetylene spectrum near 787~nm was recorded at a sample pressure of 1.5~Pa,
 and the R(9) line in the $\nu_1+3\nu_3$ band of C$_2$H$_2$
 was selected as the ``target'' line for DBT measurements.
The Gaussian width has been determined with a statistical uncertainty of 6~ppm
 through about 120 scans of the spectrum recoded in about 5 hours.
The temperatures determined from the Gaussian width indicate
 a statistical uncertainty of less than 10~ppm,
 but higher than the readings of calibrated thermal sensors by 793~ppm.
We conclude that the reason could be a result of ``hidden'' weak lines
 overlapped with the target line.
The assumption is supported by the spectrum recorded at high sample pressures,
 which reveals many lines thousands of times weaker than the known $^{12}$C$_2$H$_2$ lines in this region.
The presence of such weak lines,
 some of which could be very close to the selected target line,
 may distort the detected lineshape
 and result with an increased line width derived from the spectral fitting.

In summary, our preliminary attempt to apply a CRDS-based instrument
 for DBT studies shows that a statistical uncertainty of a few ppm
 is feasible, which makes it a promising technique to
 determine the Boltzmann constant.
The superior sensitivity of CRDS allows to detect the spectrum
 at very low sample pressures,
 which will reduce the systematic uncertainty due to incomplete knowledge on the collision-induced line profiles.
Concerning the ``hidden'' weak lines problem
 due to the complexity in the spectra of polyatomic molecules,
  lines of diatomic molecules may be more suitable for DBT measurements toward the precision of 1~ppm.
We are building a new CRDS system combined with a ring-down cavity
 thermo-stabilized at the temperature of the triple point of water.
A ro-vibrational line of CO will be used as the ``target'' line for
 the determination of $k_B$.

\section*{Acknowledgements}

The authors thank Dr. J.-T. Zhang from NIM for helpful discussions on temperature control.
This work is jointly supported by
 NSFC (91436209, 21225314 \& 91221304 ), CAS (XDB01020000)
 and NBRPC (2013BAK12B00).

\section*{References}


\providecommand{\newblock}{}


\end{CJK*}
\end{document}